\documentclass[aps,prd, preprintnumbers,groupedaddress,nofootinbib,twocolumn]{revtex4}
\usepackage{graphicx}
\usepackage{latexsym}
\usepackage{amsfonts}
\usepackage{amssymb}
\usepackage{amsmath}
\usepackage{slashed}
\usepackage{feynmp}
\usepackage{hyperref}

\pdfoutput=1

\begin{document}

%\hfill NPAC-10-NN

\title{Diagnosing Spin at the LHC via Vector Boson Fusion}

\author{Matthew R.~Buckley$^{1,2}$ and Michael J.~Ramsey-Musolf$^{3,1}$}
\affiliation{$^1$California Institute of Technology, Pasadena, CA 91125, USA}
\affiliation{$^2$Center for Particle Astrophysics, Fermi National Accelerator Laboratory, Batavia, IL 60510}
\affiliation{$^3$Department of Physics, University of Wisconsin--Madison, Madison, WI 53706, USA}
%\affiliation{$^3$Kellogg Radiation Laboratory, California Institute of Technology, Pasadena, CA 91125, USA}
\preprint{CALT 68-2801,NPAC-10-10,FERMILAB-PUB-10-342-T}
%\date{\today}

\begin{abstract}
We propose a new technique for determining the spin of new massive particles that might be discovered at the Large Hadron Collider. The method relies on pair-production of the new particles in a kinematic regime where the vector boson fusion production of color singlets is enhanced. For this regime, we show that the distribution of the leading jets as a function of their relative azimuthal angle can be used to distinguish spin-0 from spin-$\tfrac{1}{2}$ particles. We illustrate this effect by considering the particular cases of (i) strongly-interacting, stable particles and  (ii) supersymmetric particles carrying color charge. We argue that this method should be applicable in a wide range of new physics scenarios.
\end{abstract}
%%%%%%%%%%%%%%%%%%%%%%%
\maketitle

\section{Introduction \label{sec:intro}}

The TeV energy scale is expected to contain an array of new particles predicted in extensions of the Standard Model (SM) that address electroweak symmetry breaking,  naturalness, and hierarchy problems. Among the most widely discussed scenarios are supersymmetry, extra dimensions, grand unified theories, and technicolor. Assuming that new particles are discovered, the challenge will be to ascertain which new physics model -- including the possibility of one not yet considered --  best accounts for them. This will be complicated by a degeneracy among many scenarios in basic experimental signatures. A determination of the particles' quantum numbers will provide key input in discriminating among the candidate models.

One of the most important measurements -- and the focus of this paper -- is the determination of particle spin. It is, for example, one of the primary ways to distinguish supersymmetry from universal extra dimensions (see {\it e.g.}~\cite{Cheng:2002ab}).  Theorists have devoted considerable effort to developing methods for diagnosing spin at a hadron collider or a possible new linear collider \cite{Barr:2004ze,Barr:2005dz,Battaglia:2005zf,Smillie:2005ar,Wang:2006hk,Alves:2007xt,Buckley:2008pp}. 
The simplest method is to determine the kinematic dependence of the Drell-Yan pair production cross section \cite{Battaglia:2005zf}. More sophisticated methods rely on the spin-dependence encoded in the decays of the produced particles. This approach is challenging for $pp$ collisions at the Large Hadron Collider (LHC) since a full kinematic reconstruction of the decay is not possible in events with two or more invisible particles. As many extensions of the SM contain dark matter candidates that would be produced in cascades at the LHC, this is an especially limiting constraint. Consequently, one must relate the spin-dependent correlation of daughter particles in the decaying particle rest frame to an appropriate quantity in the lab frame. The ensuing loss of kinematic information renders this method less than fully general, though it can be quite powerful in certain circumstances \cite{Barr:2004ze,Barr:2005dz,Smillie:2005ar,Wang:2006hk,Alves:2007xt}. Similar considerations apply to the measurement of spin information contained in quantum interference between different decay amplitudes \cite{Buckley:2008pp}. It is therefore desirable to develop complementary spin probes free of these limitations.

%While these techniques are successful, they all suffer from the same issue: lack of kinematic information on the new particles. If, as is the case in many theories, the dark matter problem is to be solved by the new TeV-scale physics, then any new physics process will end with two invisible particles in the final state. At the LHC, this lack of information typically requires long cascade chains, which provide additional constraints, allowing the initial state kinematics to be better reconstructed. However, not all new physics scenarios will necessarily be so thoughtful as to provide such handles, and so new techniques should be sought after.

In this paper, we propose a new spin measurement technique that does not require  even partial reconstruction of the new particles' momenta.  The method relies on isolating a kinematic regime in which vector boson fusion (VBF) pair production is enhanced relative to other mechanisms ({\em e.g.}, Drell-Yan). In this regime, we focus on the associated forward (or leading) jets and show that the differential distribution as a function relative jet-jet azimuthal angle, $\Delta\phi$, can be used to diagnose the spin-statistics of the produced massive particles. In particular, we find that the sign of the $\cos 2\Delta\phi$ term in the distribution differs for spin-0 and spin-1/2 particles. The dependence on $\Delta \phi$ can be understood as the physical observable arising from the interference of helicity amplitudes associated with the vector bosons involved in the pair production of new particles. Information on the new particles' spin --  encoded in the Lorentz structure of their pair production amplitude -- is communicated to the interfering vector boson helicity amplitudes. The resulting azimuthal distribution of the leading jets probes this information. As we show below, 
by experimentally determining the sign of the $\cos2\Delta\phi$ component of the dijet distribution, one has as a result  particularly clean \lq\lq diagnostic" of the new particle's spin.

Our investigation of this technique has been heavily influenced by previous studies of dijet correlations in VBF~\cite{Eboli:2000ze,Plehn:2001nj,Hankele:2006ja,Klamke:2007cu,Hagiwara:2009wt}. The earliest work concentrated on the use of weak boson fusion (WBF) to discover an invisibly decaying Higgs boson or to probe the CP properties of the Higgs coupling to the weak bosons \cite{Eboli:2000ze,Plehn:2001nj}. At the partonic level, the WBF process can be isolated from the large QCD and $t{\bar t}$ dijet backgrounds using the distinctive pseudorapidity difference ($\Delta\eta$) and dijet invariant mass ($m_{jj}$) distributions. The resulting signal associated with the invisibly decaying Higgs involves an enhancement of the differential cross section for $\Delta\phi< \pi/2$, whereas the remaining Standard Model dijet background favors the opposite hemisphere. Subsequent Monte studies by the ATLAS collaboration indicate a persistence of this signal after accounting for parton showering and various trigger options \cite{Aad:2009wy}. A discovery of the invisibly decaying Higgs would be possible with 30 $\mathrm{fb}^{-1}$ of integrated luminosity for a wide range or Higgs boson masses if the effective cross section is more than 60\% of the SM Higgs production cross section. 

Subsequent studies of the Higgs boson CP properties considered both WBF \cite{Plehn:2001nj} and gluon fusion (GF) \cite{Hankele:2006ja,Klamke:2007cu}. The $\Delta\eta$ and $m_{jj}$ distributions for the two classes of VBF processes are quite distinct, leading to different event selection criteria for each. For both processes, however, the appearance of a distinctive $\Delta\phi$ distribution associated with various CP-even and CP-odd couplings of the Higgs to vector bosons appears to be associated with leading jets that are widely separated in pseudorapidity  (see, {\em e.g.}, Fig. 11 of Ref.~\cite{Klamke:2007cu}).  Thus, imposing an event selection cut on $|\Delta \eta|$ appears advantageous for enhancing the $\Delta\phi$ signal, even for GF for which the $\Delta\eta$ distribution itself is not useful for isolating the fusion process.

Motivated by these observations, we consider the $\Delta\phi$ distributions for pair production of new particles through the VBF process in the large $\Delta \eta$ regime. We find that as the $\Delta\eta$ cut is relaxed, the distinctive signal associated with spin-0 or spin-1/2 particles vanishes. To gain further insight into this result, we show analytically that the signal is associated with the two fusion bosons in the color (or weak isospin) singlet state. In the case of GF, color singlet dominance -- or color \lq\lq coherence" of the leading jets -- has been intuitively associated with a rapidity gap, since a color singlet object is unlikely to radiate significantly in the central region (for a discussion of these expectations, see, {\em e.g.} Ref.~\cite{Bjorken:1992rv} and references therein). Overlapping minimum bias events and minijet activity will populate  the central region -- considerations that are likely to apply to the earlier GF studies \cite{Hankele:2006ja,Klamke:2007cu} as well as in the present case. Nevertheless, 
it is intriguing that the spin-dependent signal associated with the underlying hard event is associated with both a large separation in pseudorapidity and a color singlet configuration. 

As a practical matter, one must be able to distinguish the leading jets from jets produced in the decays of the produced particle pair, particularly in the case of GF where the latter is produced strongly. In this respect, we note that
the forward jets relevant for our observable are essentially initial state radiation (ISR) associated with the hard event. Typically, ISR is an undesirable but unavoidable fact of life at hadronic colliders; ISR jets being easily confused with jets originating in decays of new physics. 
%cite
Considerable work has gone into separating these two classes of hadronic activity (see {\it e.g.}~Ref.~\cite{Alwall:2009zu,Nojiri:2010mk}, in which ISR must be distinguished from the decays of gluinos in order to make mass measurements, and Ref.~\cite{Krohn:2011zp} where jet kinematics are used to statistically identify ISR jets). These studies suggest that the prospects are promising for isolating the jets of interest to the spin determination from the decay products.  We defer a detailed study of jet identification to future work, and concentrate here on the physics of the underlying signal.%It is interesting that -- as demonstrated in this paper --  these leading jets contain valuable information on the properties of the hard event, even though they originate in SM QCD interactions.

In order to minimize complications associated with SM backgrounds, we first demonstrate the spin-dependent $\Delta\phi$ correlation  with a simulation for scalar and spinor stable strongly-interacting particles, {\it i.e.}~$R$-hadrons \cite{Farrar:1978rk,Farrar:1978xj}. This is an interesting illustrative scenario in its own right, as the spin of stable particles cannot be measured using the standard techniques (see Refs.~\cite{Allanach:2001sd,Buckley:2010fj} for exceptions). We also show, through explicit simulations, that the same effect is present for strongly interacting supersymmetric particles and for Standard Model $t{\bar t}+jj$ production.  In all these cases, the relevant vector boson is the gluon. In fact, $t{\bar t}+jj$ will be one of the primary backgrounds, as the corresponding forward jets will display the same $\Delta\phi$ distribution as for jets associated with GF production of a new pair of massive fermions. It should be possible to minimize this background using a $b$-jet veto\cite{Klamke:2007cu}, a possibility that we will study in detail along with jet identification in a forthcoming publication. We anticipate that the same spin-dependent $\Delta\phi$ signal should also be present for weak boson fusion (WBF). Although WBF can be isolated through the $\Delta\eta$ and $m_{jj}$ distributions, and backgrounds further reduced through use of the central jet veto (see, {\em e.g.}, Refs.~\cite{Barger:1990py,Barger:1991ar,Barger:1994zq,Rainwater:1999sd} and references therein),  the event rate would be very reduced. Consequently, we concentrate here on the GF case in order to demonstrate the basic concept of the technique. 

In the next Section we explain in detail the underlying concept behind our new spin measurement technique, and analytically demonstrate its utility in the toy examples of scalars and spinors charged under an abelian gauge group. In Section~\ref{sec:nonabelian}, we apply the technique to simples models with non-abelian QCD interactions, using MadGraph/MadEvent \cite{Alwall:2007st} and CalcHEP simulations \cite{Pukhov:1999gg}. As will be shown, inclusion of color factors greatly complicates the analytic calculation, but does not change the overall applicability of the method.

\section{Analytic Calculations with Abelian Examples \label{sec:abelian}}

As outlined in the Introduction, we are interested in the pair production via VBF of two heavy particles with two forward jets. The interference of the vector bosons helicities is set by the Lorentz structure of the pair production matrix element, and in turn affects the distribution in $\Delta\phi$ of the two forward jets. To make this causal chain clear, we begin with an abelian gauge group as the vector bosons which fuse, and consider only the dominate Feynman diagrams in order to make the computation tractable. The full calculation of abelian gauge groups, kinematic cuts, and all involved diagrams requires simulation to perform, and will be described in Section~\ref{sec:nonabelian}.

\begin{figure}[t]
\includegraphics[width=\columnwidth]{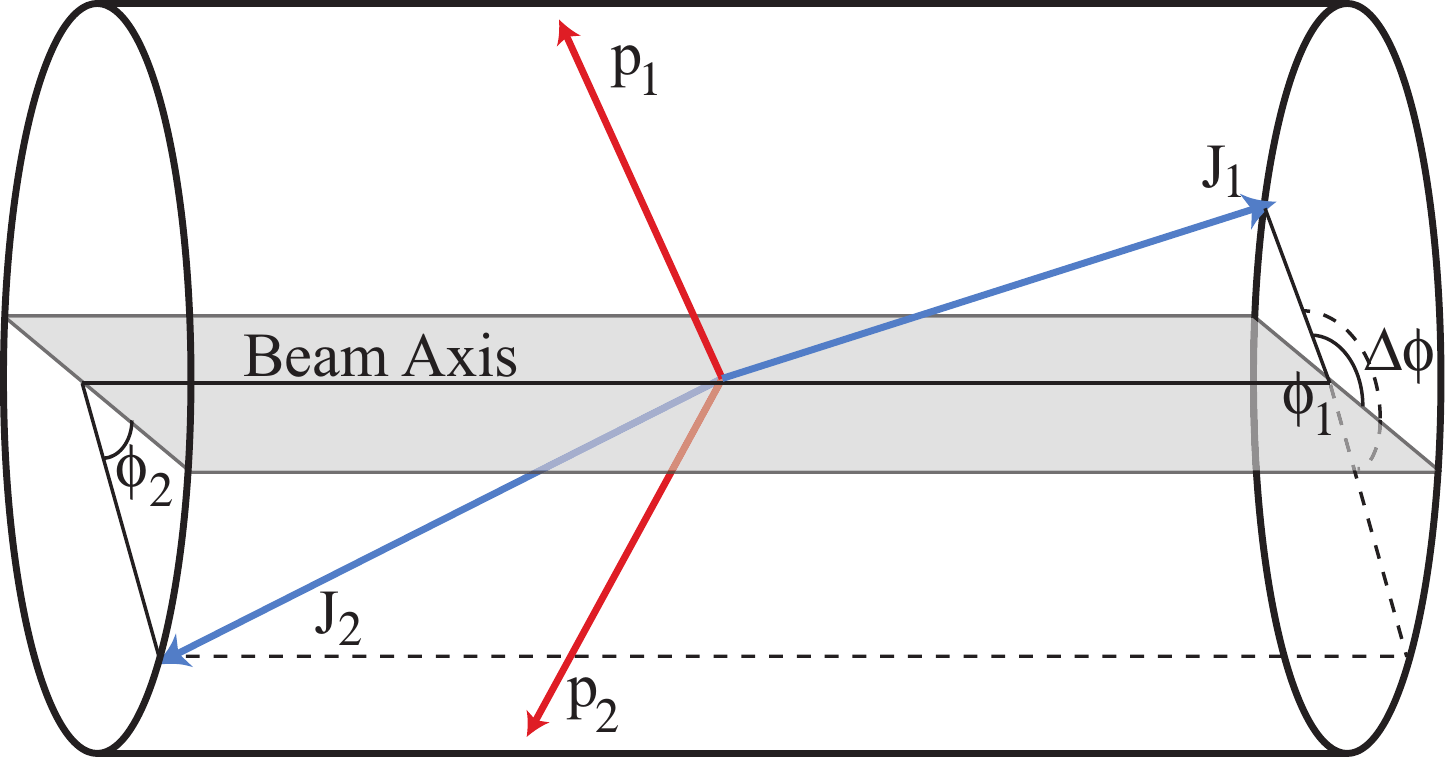}
\caption{A VBF event in an idealized detector volume.  \label{fig:kinematics}}
\end{figure}

To set the theoretical framework, we consider pair production of massive particles in a regime where VBF is kinematically favored. As illustrated in Fig.~\ref{fig:kinematics}, the event contains two forward jets ($J_1$ and $J_2$) as well as the two new particles (each having mass $m$ and momenta $p_1$ and $p_2$). The azimuthal angle $\phi_j$ of each jet is defined relative to a reference plane. The location of the reference plane is arbitrary, as is direction of the $z$-axis, making the labeling of jets as \lq\lq 1" and \lq\lq 2" arbitrary as well (in the absence of jet tagging). Consequently, the only physically quantity is the relative azimuthal angle $\Delta\phi$.

\begin{figure}[t]
\includegraphics[width=\columnwidth]{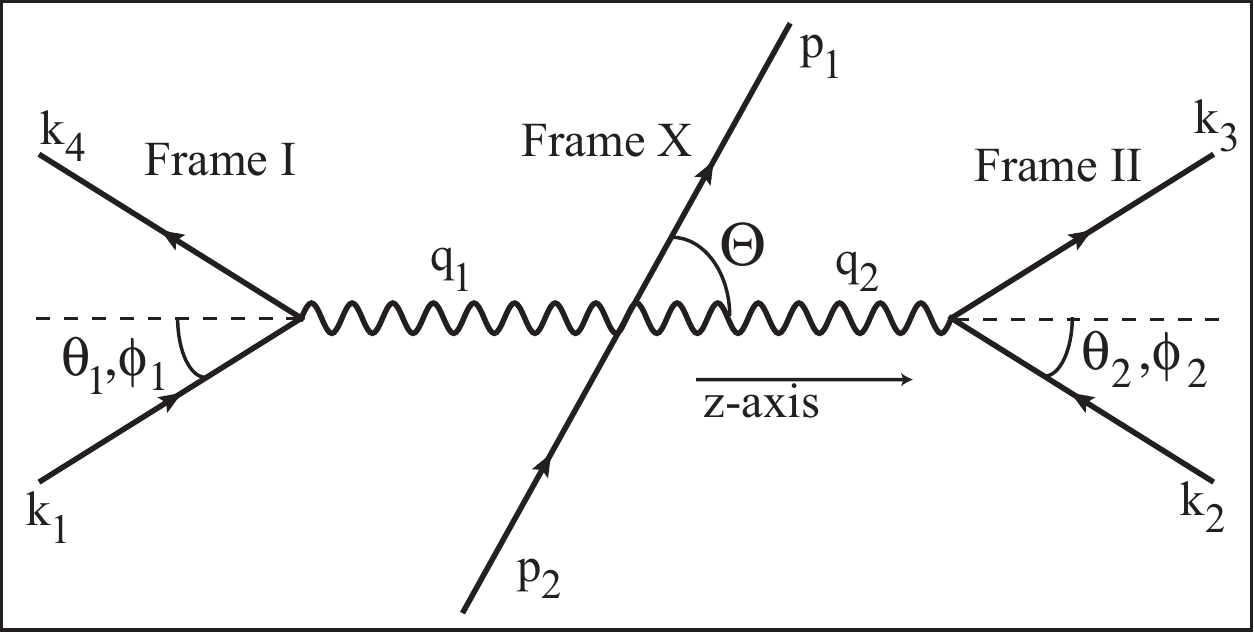}
\caption{Kinematics for VBF in Frames I, II and X. Azimuthal angles $\phi_1$ and $\phi_2$ are defined relative to the production plane of the massive particles (momenta $p_1$ and $p_2$). Polar angles $\theta_1$, $\theta_2$ and $\Theta$ are defined relative to the beam axis in their respective frames. \label{fig:VBFkinematics}}
\end{figure}

Our labeling of the kinematics of the VBF amplitude is indicated in Fig.~\ref{fig:VBFkinematics}, largely following the conventions of Ref.~\cite{Hagiwara:2009wt}. The polarization vectors of the gauge bosons are proportional to $e^{ih_i\phi_i}$, where $i=1,2$, and $h_i$ is the polarization of the boson defined relative to the positive $z$-axis. The matrix elements for the emissions of the gauge bosons can be expressed as 
$
{\cal M}_i(\theta_i,\phi_i,h_i) = e^{ih_i\phi_i}{\cal M}_i(\theta_i,0,h_i). 
$
Thus, any dependence on $\phi_1$ and $\phi_2$ in the amplitude must be due to interference between matrix elements of the different polarizations. The resulting contribution to the differential cross section then yields set of 24 functions (sines and cosines) whose arguments are linear combinations of the $\phi_j$. As discussed above, in this paper we retain only the terms dependent on $\Delta\phi$. Moreover, if the pair production amplitude is CP-invariant as we will assume here, then only the cosine terms survive. Thus, 
\begin{equation}
\frac{d\sigma}{d\Delta\phi}  =  A_0 +A_1 \cos \Delta \phi +A_3\cos2\Delta \phi 
 \ \ . \label{eq:deltaexpansion}
\end{equation}

%Barring some way to experimentally access the production plane formed by the two massive particles, we can only measure $\Delta \phi \equiv \phi_1-\phi_2$. While this may be possible in some cases (such as the completely reconstructable $R$-hadrons in Section~\ref{sec:rhadrons}), in this paper we consider only the simplest case and reparametrize the angles as $\Delta \phi$ and $\Sigma = \phi_1+\phi_2$. Then most general possible dependence on $\Delta \phi$ for the differential cross section with respect to this angle is

The VBF kinematics induces a basic $\cos\Delta\phi$ dependence on the cross section that typically overshadows the contribution to $A_1$ from the interfering amplitudes \cite{Eboli:2000ze}. The fusion process tends to knock the produced pair and recoiling jets on opposite sides of a suitably defined reference plane, favoring a smaller $\Delta\phi$. In resonant production, this $\Delta\phi$ bias can be exploited in searches for \lq\lq invisibly" decaying Higgs bosons produced through WBF \cite{Eboli:2000ze,Aad:2009wy}. In contrast, the $\cos2\Delta\phi$ term is dominated by the interference of the VBF amplitudes, receiving no significant kinematic contribution. Importantly, we find that the sign of the coefficient function $A_3$ is opposite for scalars (positive $A_3$) and spinors (negative $A_3$). While in this Section, we only demonstrate the sign flip between the two spin choices, the results of our simulations in Section~\ref{sec:nonabelian} will show that, while the $\cos2\Delta\phi$ correlation is subleading, it is sufficiently large to make its relative sign a useful spin diagnostic.

%Before discussing these simulations in detail, it is instructive to develop some intuition for the origin of the spin-dependent sign difference.  
The coefficient $A_3$ arises from the interference of the VBF pair production matrix elements ${\cal M}_{\rm pair}$ that differ by two unit of boson polarization:
%\begin{widetext}
\begin{eqnarray}
\nonumber
A_3 &=& ({\cal PS})  \sum_{{h_1,h_1',h_2,h_2'}\atop{|h_i-h_i'|=2}} \Bigl({\cal M}_1(h_1){\cal M}_2(h_2){\cal M}_{\rm pair}(h_1,h_2) \Bigr)\\
&&\Bigl({\cal M}_1(h_1'){\cal M}_2(h_2'){\cal M}_{\rm pair}(h_1',h_2') \Bigr)^\ast \ \ \ . \label{eq:f9def}
\end{eqnarray}
%\end{widetext}
Here $({\cal PS})$ is the phase space factor;  ${\cal M}_1$ and ${\cal M}_2$ are the parton $\to$ parton + gauge boson
production amplitudes (see Fig.~\ref{fig:VBFkinematics}); and ${\cal M}_{\rm pair}$ is the VBF amplitude for production of the heavy pair. Performing the helicity sums leads to 
%\begin{widetext}
\begin{eqnarray}
A_3  &=& ({\cal PS}){\cal M}_1(+1){\cal M}_1(-1)^*{\cal M}_2(-1){\cal M}_2(+1)^* \\
\nonumber
&\times& {\cal M}_{\rm pair}(+1,-1){\cal M}_{\rm pair}(-1,+1)^*+(1\leftrightarrow -1)
 . \label{eq:f9explicit}
\end{eqnarray}
%\end{widetext}

In the illustrative simple case considered in this Section, that of abelian gauge bosons ({\em e.g.}, photons), the fusion amplitude ${\cal M}_{\rm pair}$ is generated by the standard Compton diagrams (with gauge coupling $g$). For spin-0 particles of charge $Q$, a straightforward computation gives
%\begin{widetext}
\begin{eqnarray}
\nonumber
{\cal M}_{{\rm scalar}} & = & g^2 Q^2\Bigl\{ (\epsilon_1\cdot\epsilon_2)
-4  \Bigl[ \frac{(p_1 \cdot \epsilon_1)(p_1-q_1)\cdot \epsilon_2}{q_1^2-2p_1\cdot q_1}\\
&&+\frac{ (p_1 \cdot \epsilon_2)(p_1-q_2)\cdot \epsilon_1 }{q_2^2-2p_1\cdot q_2}\Bigr]\Bigr\}. 
\label{eq:scalarmatrix}
\end{eqnarray}
%\end{widetext}
Choosing polarization vectors (with $h_i = -1,0,+1$) to reflect the angular momentum in the $z$-direction, $\epsilon_1^\pm = \epsilon_2^\pm \equiv \epsilon^\pm$. 
%Then, ${\cal M}_{\rm pair}$ is due to four diagrams: an $s$-channel process via a gauge boson, a four-point interaction involving two scalars and two vectors, and two $t$-channel diagrams. The resulting matrix element for the scalars is
%\begin{widetext}
%\begin{eqnarray}
%{\cal M}_{{\rm scalar}} & = & g_s^2 \left\{ f^{abc}T^c \frac{(\epsilon_1\cdot\epsilon_2)(q_1- q_2)\cdot(p_1 - p_2)+2 (q_2 \cdot \epsilon_1)(p_1-p_2)\cdot \epsilon_2-2 (q_1 \cdot \epsilon_2)(p_1-p_2)\cdot \epsilon_1 }{(q_1+q_2)^2} \right. \nonumber \\
% & & \left.+i\{T^a,T^b\} (\epsilon_1\cdot\epsilon_2)-4 i \left[ \frac{T^aT^b(p_1 \cdot \epsilon_1)(p_1-q_1)\cdot \epsilon_2}{q_1^2-2p_1\cdot q_1}+\frac{T^bT^a (p_1 \cdot \epsilon_2)(p_1-q_2)\cdot \epsilon_1 }{q_2^2-2p_1\cdot q_2}\right]\right\}. \label{eq:scalarmatrix}
%\end{eqnarray}
%\end{widetext}
It is then obvious that ${\cal M}(+1,-1) = {\cal M}(-1,+1)$, which demonstrates that $A_3 \propto |{\cal M}_{\rm scalar} (\pm1,\mp1)|^2$, and is therefore positive. Though a more lengthly calculation is involved, this argument also holds for VBF scalar pair production by non-abelian gauge bosons.

The calculation for spinors is similarly straightforward, but the origin of the negative sign of $A_3$ (opposite that of the scalars) is less transparent.  However, a closed form expression can be obtained in the limit that the two abelian gauge bosons are 
are massless and on-shell ($q_1^2=q_1^2= 0$). Defining $\lambda=\sqrt{1-4m^2/s}$ where $s=(k_1+k_2)^2$, we find
\begin{equation}
{A_3} \propto - \frac{64m^2}{s}\left(1+\frac{4m^2}{s\lambda}\tanh^{-1}\lambda \right), \label{eq:abelianf9}
\end{equation}
which is clearly negative. It is interesting to note that the minus sign (relative to the scalar calculation), arises from the anticommutation of fermion operators in the simplification of the expression for ${\cal M}_{\rm pair}(+1,-1){\cal M}_{\rm pair}(-1,+1)^*$, thus making the connection between the sign of $\cos2\Delta\phi$ and spin clear. 

We now briefly mention the difficulties in analytic calculation for the non-abelian fermionic example, before continuing on to the numeric simulations. In the explicit calculation of the non-abelian example, two terms arise: a positive contribution with an antisymmetric color factor $[T^a,T^b]$, and a negative contribution -- identical to Eq.~(\ref{eq:abelianf9}) up to an overall a symmetric color factor $\{T^a,T^b\}$. These two terms correspond to the two gluons being in a color octet and color singlet state, respectively. As we shall see, numeric calculation for off-shell gauge bosons confirms that the coefficient is overall negative, meaning that the color singlet state (symmetric term) dominates. This is an intriguing result, and we give a heuristic explanation.
Helicity conservation and the kinematics of nearly forward jets (typical of ISR) favors a spatially symmetric configuration for the fusing  gluons, and thus a color symmetric (singlet) two gluon state. However, without an application of kinematic cuts on the 4-body phase space, this is not reflected in the analytic calculation. 

%We begin with an abelian gauge group; in which case the only pair production diagrams involved are the two $t$-channel diagrams (as the $s$-channel diagram vanishes without a non-abelian three-point interaction). The pair production matrix element can be simplified to
%\begin{widetext}
%\begin{eqnarray}
%{\cal M}_{\rm spinor}(\pm1, \mp1) = -g^2 \bar{u}(p_1)\left[\frac{2 p_1\cdot \epsilon^\pm \slashed{\epsilon}^\mp + \slashed{q}_1\slashed{\epsilon}^\pm \slashed{\epsilon}^\mp}{q_1^2-2p_1\cdot q_1} +\frac{2 p_1\cdot \epsilon^\mp \slashed{\epsilon}^\pm + \slashed{q}_2\slashed{\epsilon}^\mp \slashed{\epsilon}^\pm}{q_2^2-2p_1\cdot q_2} \right]v(p_2).
%\end{eqnarray}
%\end{widetext}
%Using this matrix element in Eq.~(\ref{eq:f9explicit}), it is not immediately clear that the resulting coefficient for $A_3$ is negative. Explicit calculation is possible; though the end result does not yield a compact closed form solution. 

%In the limit that the two gauge bosons are massless and on-shell ($q_1^2=q_1^2= 0$), we can integrate over the pair production angle $\Theta$ and find that
%\begin{equation}
%A_3 \propto - \frac{64m^2}{s}\left(1+\frac{4m^2}{s\sqrt{1-\frac{4m^2}{s}}}\tanh^{-1}\sqrt{1-\frac{4m^2}{s}} \right), \label{eq:abelianf9}
%\end{equation}
%which is clearly negative. Numeric calculation of off-shell gauge bosons confirms that the coefficient is negative in those limits as well.

\section{Simulations with QCD Gauge Forces \label{sec:nonabelian}}

We therefore turn to a full calculation of the jet distributions using non-abelian gauge boson. As outlined previously, a realistic computation involves additional subtleties, though we find from simulations that the sign difference persists. The non-abelian Compton amplitude ${\cal M}_{\rm pair}$ contains a dependence on the structure constants and generators of the gauge group as well as contributions from a diagram involving fusion of two gauge bosons to a third, followed by production of the heavy pair. In addition, the coefficients $A_0$, $A_1$ and  $A_3$ receive additional contributions from heavy particle pair production via a bremsstrahlung gluon. If production was primarily via an on-shell intermediary ({\em i.e.}, a resonance), the $\Delta \phi$ signal would be governed by the coupling of the intermediary to the two gauge bosons rather than by the spin of the produced pair. However, such production is not dominate in the cases we consider here.

%The case of non-abelian gauge groups is even more complicated. Here we can split the matrix elements into terms with symmetric combinations of the initial gluon colors $\{T^a,T^b\}$, and terms with antisymmetric combinations $[T^a,T^b]$. The former tends to contribute negatively to the total $A_3$, and the latter makes positive contributions. However, this ignores interference between graphs that have other color structures. A full calculation must be performed, including all these terms, in order to see whether the overall coefficient is positive or negative. In the next section, we use automated methods to perform just such a calculation. The results match the abelian case: the spinor $A_3$ is negative.

%Though the end result, that the coefficient for scalars is positive and for negative for spinors, seems somewhat arbitrary, it should be noted that the different Lorentz structure of the two interactions are at the root of the different signs. The scalar matrix element depends only on dot products of the polarization and momentum vectors; so the order in which $\epsilon^+$ and $\epsilon^-$ enter into the calculation does not matter. With the introduction of non-commutating gamma matrices, the spinor matrix elements can (and indeed do) have different signs for ${\cal M}(+1,-1)$ and ${\cal M}(-1,+1)$. This provides a clear-cut method to differentiate the two spin cases, as we will show with simulations in the next section.

%\section{$R$-Hadron Spin Measurement \label{sec:rhadrons}}

To illustrate the spin-dependent sign of the $\cos 2\Delta\phi$, we first specialize to VBF production of $R$-hadron pairs  at the LHC (for an overview and references, see {\em e.g.}, Ref.~\cite{Fairbairn:2006gg}). These stable, strongly interacting particles would have striking signatures at the LHC, either appearing as ``slow muons'' that are highly ionizing \cite{Drees:1990yw,Fairbairn:2006gg}, or as stopped tracks in the detector \cite{atlassearch,Arvanitaki:2005nq}. The background is therefore limited to detector errors, which we will ignore for this study.  $R$-hadrons were originally proposed in supersymmetric models with a gluino or squark as the lightest (or next-to-lightest) supersymmetric particle \cite{Baer:1998pg,Mafi:1999dg,Mafi:2000kg,Raby:1997bpa,Raby:1997pb,Dreiner:1997uz,Berger:2003kc}, though other theories can provide similar particles (for an excellent overview, see Ref.~\cite{Fairbairn:2006gg}). 
Null results from searches from ALEPH \cite{Heister:2002hp}, CDF \cite{Acosta:2003ys}, and LEP2 \cite{SUSYwg}, exclude particles of masses less than $\sim 200-250$~GeV, depending on the theoretical assumptions made, while the LHC should be able to find $R$-hadrons up to a few TeV in mass \cite{Fairbairn:2006gg,Mermod:2009ct}. In addition, since standard spin measurements cannot be applied to stable particles, new techniques would be necessary in order to study the spin of $R$-hadrons \cite{Buckley:2010fj}.

We assume some new ${\mathbb Z}_2$ quantum number ($R$-parity in supersymmetry) that requires $R$-hadrons to be pair produced. As strongly interacting particles, this production can proceed via VBF mediated by gluons (\lq\lq gluon fusion,"  or ``GF"). Matrix elements for two heavy fermions or scalars plus two jets from $pp$ initial states are generated in MadGraph, and Monte Carlo simulations in MadEvent \cite{Alwall:2007st}; such simulations correctly include all interference terms as well as production from a bremsstrahlung gluon. To isolate the kinematics in which the GF process is enhanced relative to Drell-Yan, we impose the following cuts:
\begin{eqnarray}
\eta_{j_1} \cdot\eta_{j_2} < 0,~  |\eta_{j}| \leq 5,~  |\eta_{j_1}-\eta_{j_2}|  \geq  4.2 \nonumber \\
p_{T,j_1}  \geq  30 ~\mbox{GeV},~ p_{T,j} \geq 20 ~\mbox{GeV},~ M_{jj} \geq 500 ~\mbox{GeV} \label{eq:jetcuts} \\
|\eta_{R-{\rm hadron}}| < 2.1, ~  ~p_{T,R-{\rm hadron}} > 50~\mbox{GeV}. \nonumber 
\end{eqnarray}
Here, $\eta_j$ is the jet pseudo-rapiditiy; $M_{jj}$ is the dijet invariant mass; and $p_T$ denotes the transverse momentum of a given particle.  
The jet cuts are necessary in order to reduce the non-VBF events \cite{Ciccolini:2007jr,Ciccolini:2007ec}, while the $R$-hadrons cuts are of lesser importance, and simply require the heavy states appear in the barrel of the detectors. 

\begin{figure}[t]
\includegraphics[width=\columnwidth]{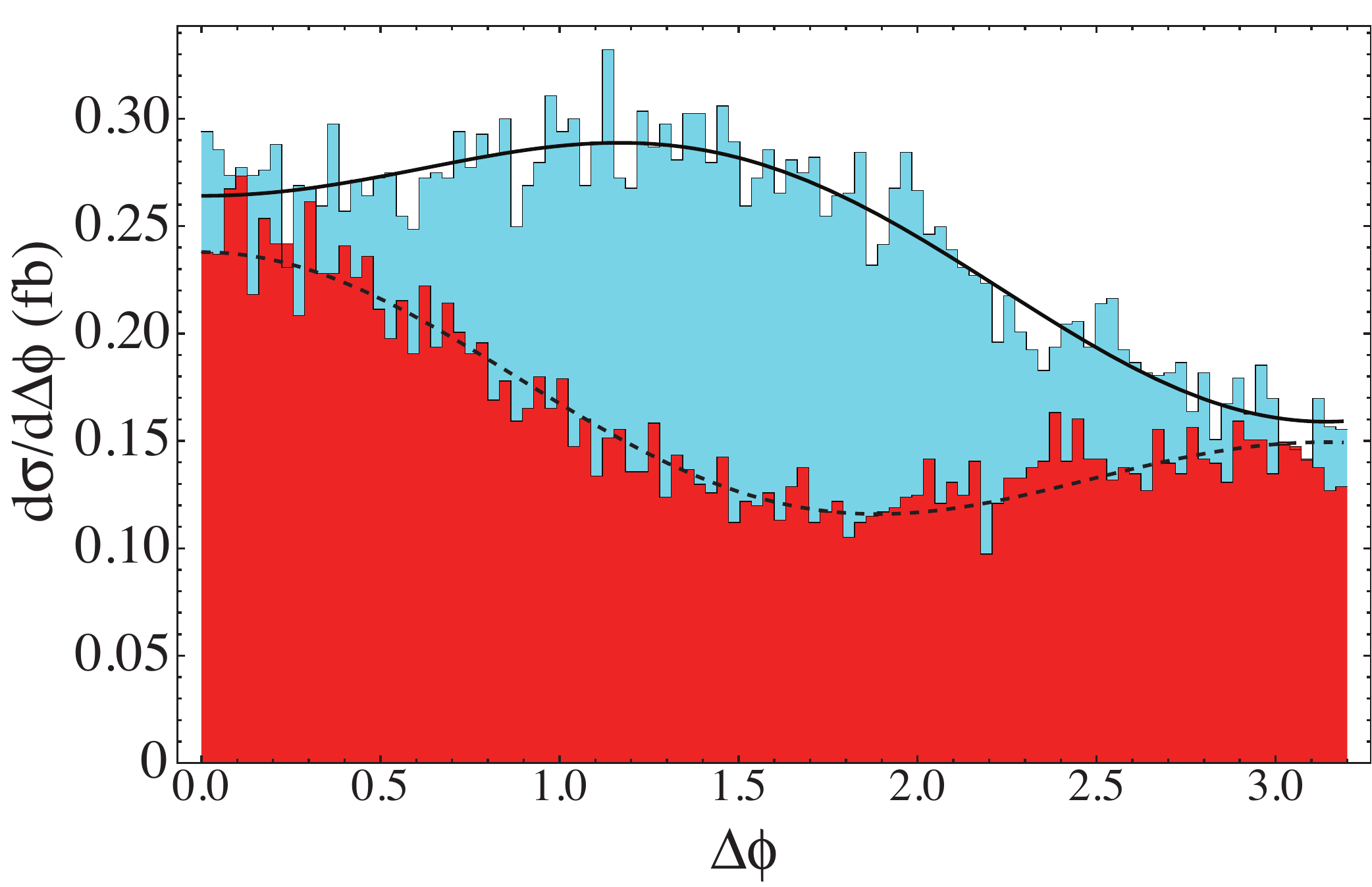}
\caption{Differential cross section $d\sigma/d\Delta\phi$ as a function of difference in the azimuthal angle of VBF jets $\Delta \phi$. The differential cross section for fermionic R-hadron pair production is shown in blue, and the best fit to $A_0 + A_1\cos\Delta\phi+A_3\cos2\Delta\phi$ is shown with a solid line. The scalar cross section is in red, and the best fit is shown with a dotted line. \label{fig:diffcross}}
\end{figure}

We generate 40,000 events for the LHC at $\sqrt{s}=10$~TeV for simple, 500~GeV, scalar and Dirac spinor $R$-hadrons, in which the heavy particles only couple to gluons. After all cuts the spinor cross sections are 33~fb (spinor) and  21~fb (scalar). The differential cross sections $d\sigma/d\Delta\phi$ for the two models are shown in Fig.~\ref{fig:diffcross}, along with the best fits to Eq.~(\ref{eq:deltaexpansion}).

It has been demonstrated that the VBF-isolating jet cuts outlined in Eq.~(\ref{eq:jetcuts}) tend to induce a positive $\cos\Delta \phi$ mode in the distribution \cite{Eboli:2000ze}. The expected positive $A_1$ mode is found both the scalar and spinor simulations. The $\cos2\Delta\phi$ coefficient $A_3$ is not affected by the cuts, and we see a significant difference between the two spin cases. The scalar case, as predicted, has a positive coefficient, while the spinor has a negative one. Appropriately normalized to the constant term, we find the scalar $A_3/A_0$ is $0.22$, while in the spinor example, $A_3/A_0=-0.14$. Without background, the statistical significance can be estimated as $\sim \left| A_3/A_0 \right| \sqrt{N}$, necessitating $\sim 30$~fb$^{-1}$ of luminosity for measurement at several sigma.  We find that the dominant partonic subprocess involves gluon radiation from light quarks.  We also observe that the $\Delta\phi$ signal disappears when the pseudo-rapidity cuts are removed, indicating the dependence of this method on kinematically selecting a region in which the color-singlet VBF process is relatively enhanced.

\begin{figure}[t]
\includegraphics[width=\columnwidth]{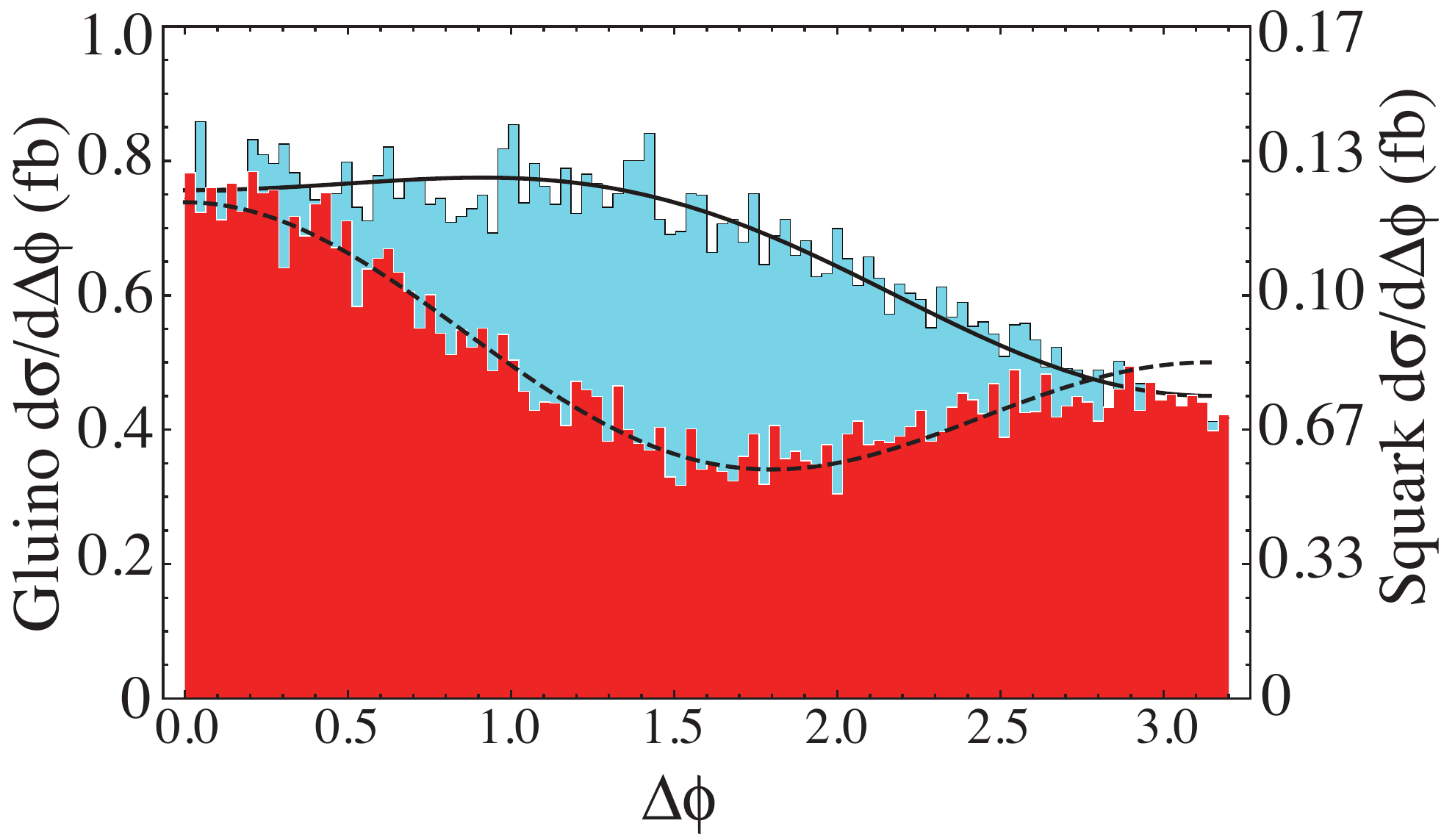}
\caption{Differential cross section $d\sigma/d\Delta\phi$ as a function of $\Delta \phi$. The differential cross section for gluino pair production is shown in blue (scale on left), and the best fit with a solid line. The right-handed up-type squark cross section is in red (scale to right), and the best fit is shown with a dotted line. \label{fig:diffcrosssusy}}
\end{figure}

The correlation between spin and the sign of $A_3$ is independent of the charge conjugation of the produced fermions. In the case of supersymmetry, for example, one may wish to determine the spin of strongly interacting superpartners (squarks and gluinos) using GF. In this case the gluinos are Majorana fermions, in contrast to the Dirac R-hadron fermions explored above. The results of an analogous simulation for 550 GeV up-type squarks and 600 GeV gluinos are given in Fig.~\ref{fig:diffcrosssusy}. Here we again see the sign change of the $\cos 2\Delta\phi$ term when comparing squark ($A_3/A_0 = 0.24$) and gluino ($A_3/A_0 = -0.09$) pair production. In contrast to the $R$-hadron case, isolating the signal associated with pair production of superpartners will require additional cuts to suppress Standard Model backgrounds. As our focus here is on the basic signal rather than on its isolation, we defer a detailed study of background reduction to a future analysis.

To ensure that the effect is not an artifact of our choice of event generator, we have repeated the calculation for the dominant partonic sub-process using Calchep \cite{Pukhov:1999gg} and confirmed the MadGraph/MadEvent results. As a final check we have computed Standard Model $t{\bar t}+jj$ production in GF kinematics and find a negative sign for $A_3$.  Depending on the heavy particle mass $m$, $t{\bar t}+jj$ could thus become a significant background that would mimic the $\Delta\phi$ signal associated with  new fermions. Implementing a $b$-jet veto may be used to reduce this background \cite{Hankele:2006ja,Klamke:2007cu}. 

\section{Conclusion \label{sec:conclusion}}

In this work, we have demonstrated a proof of principle for the use of the sub-leading $\Delta\phi$ distribution as a spin diagnostic in a kinematic regime where color-singlet GF appears to be kinematically enhanced and where the heavy particle pair production is dominated by their lowest order gauge interactions. We anticipate that the method will generalize to higher spin states and to pair production through weak vector boson fusion. Strong production of $R$-hadrons provides the cleanest illustration as the signal has a large cross section and is essentially background free. Other cases -- such as VBF production of supersymmetric particles --  in which the heavy particle decays produce additional jets and/or missing energy will require additional cuts for appropriate signal isolation. As indicated above, recent investigations in this direction \cite{Alwall:2009zu,Nojiri:2010mk} are encouraging for the prospects of appropriate jet identification in the context of GF.  
A detailed study of these issues, background suppression, and other practical considerations will appear in forthcoming work.

\section*{Acknowledgements}
We would like to thank the Aspen Center for Physics where a substantial portion of this work was carried out. We also thank J.~Alwall, T.~Han, M.~Herndon, B.~Mellado, D.~Morrissey, T.~Plehn, W.~Smith, M.~Spiropulu, J.~Thaler, and L.-T.~Wang for helpful discussions and N. Christensen for assistance with the Calchep package. This work was supported in part under Department of Energy contracts DE-FG03-92-ER40701 (MRB) and DE-FG02-08ER41531(MJRM) and by the Wisconsin Alumni Research Foundation (MRJM).

\bibliography{vbfpaper}
\bibliographystyle{apsrev}

\end{document}